\documentclass{article}
\usepackage{spconf,amsmath,amssymb,graphicx,booktabs,times}
\usepackage[hidelinks]{hyperref}
\title{AISHELL8-FISHEYE: A Fisheye Audio-Visual Dataset for Target Speaker Extraction with Distortion-Aware Baselines}
\name{\normalsize Peijun Yang$^1$, Zhan Jin$^3$, Juan Liu$^{2,*}$, Hui Bu$^4$, Xiaoyi Qin$^5$, Ruyi Gan$^5$, Hao Wang$^5$, Ming Li$^{2,3,*}$\thanks{$^*$Corresponding authors: Juan Liu (liujuan@whu.edu.cn) and Ming Li (mingli369@cuhk.edu.cn).}}
\address{$^1$School of Cyber Science and Engineering, Wuhan University, Wuhan, China\\
$^2$School of Artificial Intelligence, Wuhan University, Wuhan, China\\
$^3$School of Artificial Intelligence, The Chinese University of Hong Kong, Shenzhen, China\\
$^4$Beijing AISHELL Technology Co., Ltd., China\\
$^5$X Square Robot, Shenzhen, China}
\begin{document}
\maketitle
\begin{abstract}
Fisheye cameras offer wide visual coverage but distort facial and lip geometry, challenging audio-visual target speaker extraction. We introduce AISHELL8-FISHEYE, a 47.03-hour corpus of 94 Mandarin speakers with synchronized audio and paired fisheye and normal-camera videos. Speaker-disjoint splits and synthetic two-speaker mixtures support reproducible evaluation under matched acoustic conditions. We also establish a baseline that combines a lip-reading encoder with a frozen distortion-aware branch transferred from fisheye head pose estimation. On fisheye video, adding this branch to an adapted lip encoder improves scale-invariant signal-to-distortion ratio improvement from 10.47 to 11.84 dB. The branch does not improve performance with a frozen lip encoder, highlighting the importance of visual adaptation. A substantial gap to paired normal-camera inputs remains. The corpus and baseline resources support further research on speech extraction from distorted visual cues.
\end{abstract}

\section{Introduction}
\begin{figure}[t]
\centering
\includegraphics[width=0.80\linewidth]{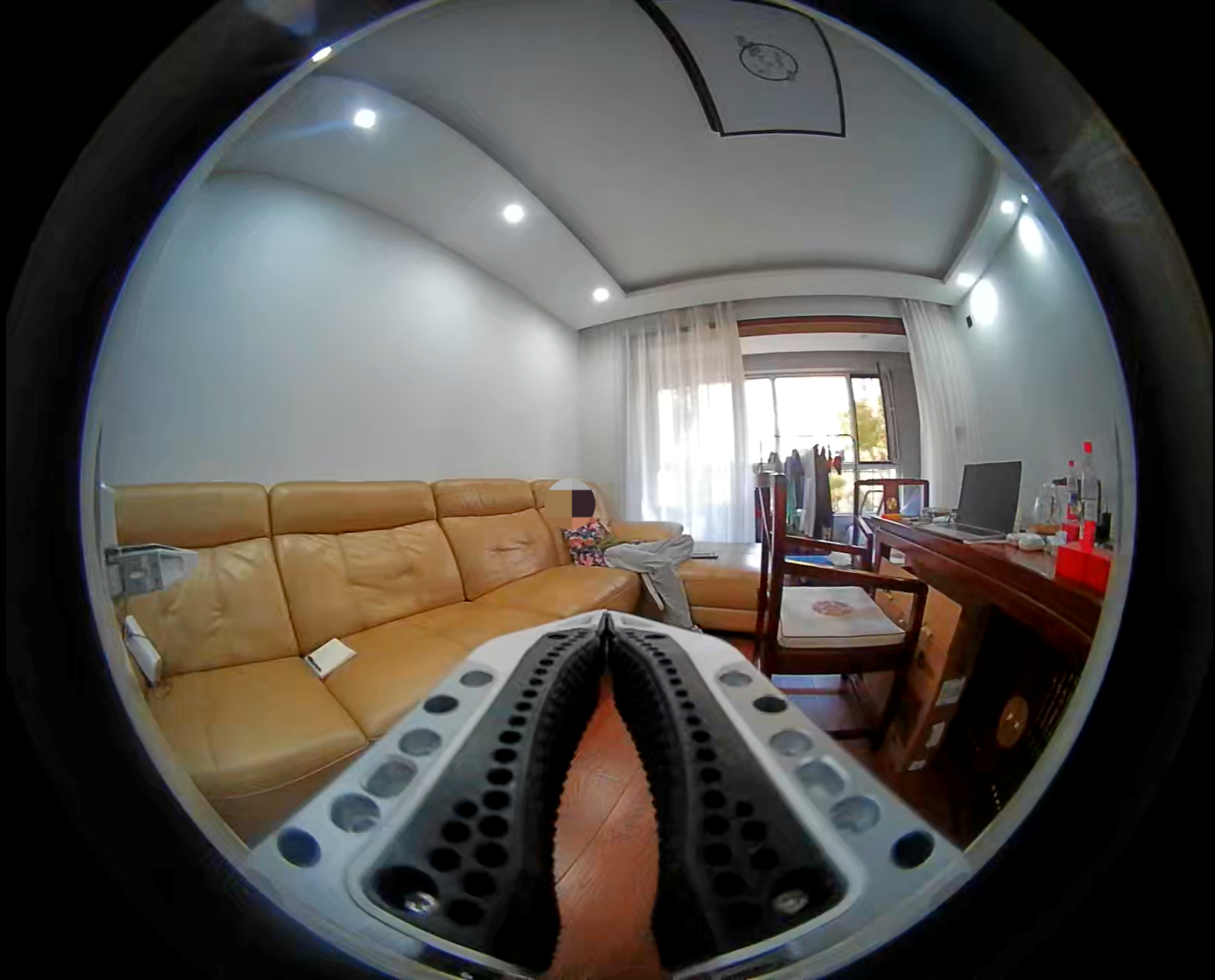}
\caption{Real fisheye view from a robot-arm-mounted camera, showing wide scene coverage and radial distortion.}
\label{fig:robot}
\end{figure}
Target speaker extraction (TSE) recovers a desired voice from overlapping speech using auxiliary information. Audio-visual TSE (AVTSE) exploits synchronized lip motion to identify the target without requiring a separate enrollment utterance. Prior work has developed visual conditioning and multimodal fusion strategies~\cite{afouras2018conversation,pan2021muse,li2023iianet}, including methods for target absence and sparse overlap~\cite{pan2022usev}. These systems motivate the use of video when acoustic interference obscures the target speech.

Fisheye cameras provide broad coverage for wearable devices and embodied robots, as well as surveillance, augmented/virtual reality, and smart conference systems~\cite{huang2024active}. Figure~\ref{fig:robot} shows a real image from a robot-arm-mounted fisheye camera, capturing the gripper and surrounding room in one view. For robots, wide-angle views can support navigation, obstacle monitoring, and tracking nearby interlocutors; lip-guided extraction could help recover a user's speech during multi-person interactions. However, radial distortion changes facial geometry across the image plane~\cite{fan2022wide}, so lip representations learned from conventional video may transfer poorly. Image rectification~\cite{fan2022wide} is a potential remedy, although it requires camera calibration and changes the sampled image through resampling. Feature adaptation~\cite{zhang2025distortion} provides another direction that can be studied directly on unrectified fisheye video.

The challenge concerns both the appearance of individual mouth shapes and the consistency of visual cues over time. Speech-related motion must remain useful after ROI detection and alignment, even when the source image has spatially varying geometry. A pretrained lip-reading network supplies articulatory information, whereas features transferred from fisheye vision may provide complementary spatial context. We therefore test whether adding fisheye spatial features improves extraction and whether the lip encoder needs to be fine-tuned to use them effectively.

Progress on this problem requires synchronized recordings with clean speech references and a reproducible evaluation protocol. We introduce AISHELL8-FISHEYE, a corpus of spontaneous Mandarin monologues recorded with fisheye and normal cameras. Its paired videos support comparisons using identical audio mixtures, while speaker-disjoint splits enable evaluation on unseen speakers. We further adapt a distortion-aware (DA) branch from fisheye head pose estimation~\cite{zhang2025distortion} to a lip-conditioned TF-GridNet separator~\cite{wang2023tf}. Our contribution is the dataset and its AVTSE benchmark; the DA branch is a transferred baseline component.

Experiments show that DA features improve extraction when the lip encoder is adapted to fisheye inputs, but provide no gain when it is frozen. We report this interaction alongside the remaining normal/fisheye performance gap. Because the cameras differ in viewpoint and resolution, the paired comparison measures the effect of the visual acquisition conditions rather than isolating radial distortion alone.

\section{AISHELL8-FISHEYE}
\subsection{Acquisition and paired recordings}
The corpus contains 94 native Mandarin speakers recorded individually in a controlled studio. Each speaker contributes approximately 30 minutes of spontaneous monologues about daily life, personal experiences, and hobbies. The recordings retain natural pauses, fillers, and topic changes. Table~\ref{tab:data} summarizes the 47.03 hours of speech and 14,791 utterances. Speakers do not overlap across training, validation, and test sets.

\begin{table}[t]
\centering
\caption{AISHELL8-FISHEYE statistics. Splits are speaker-disjoint.}
\label{tab:data}
\begin{tabular}{lrrr}
\toprule
Split & Speakers & Utterances & Hours\\
\midrule
Training & 66 & 10,308 & 32.75\\
Validation & 14 & 2,213 & 7.22\\
Test & 14 & 2,270 & 7.06\\
\midrule
Total & 94 & 14,791 & 47.03\\
\bottomrule
\end{tabular}
\end{table}

\begin{figure}[t]
\centering
\includegraphics[width=\linewidth]{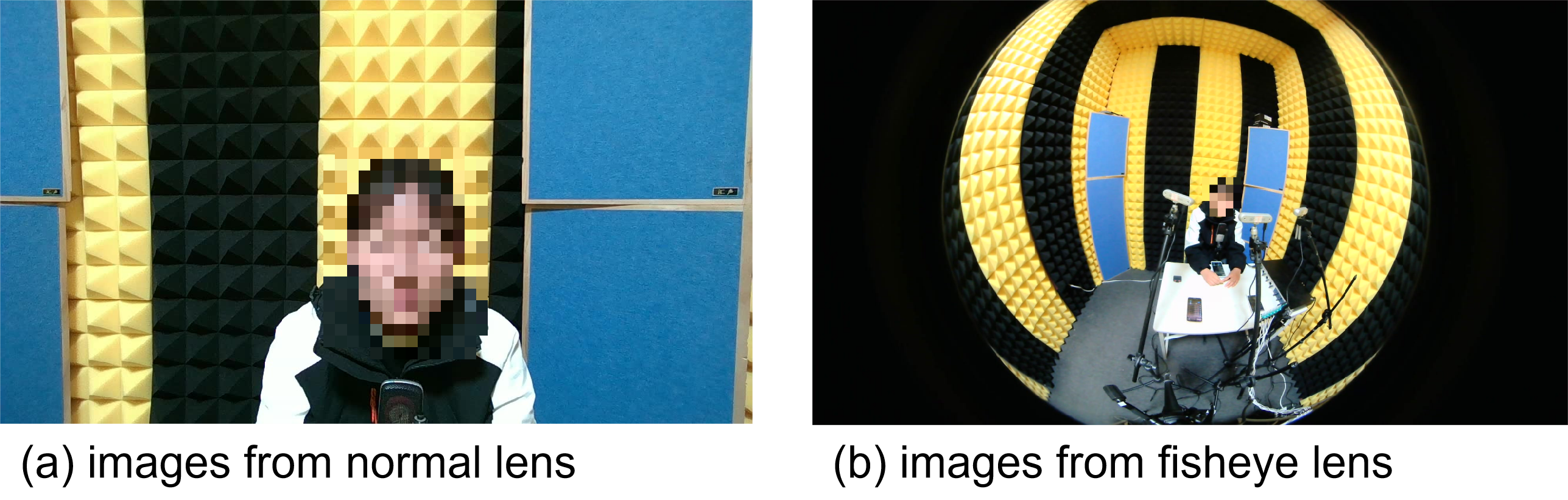}
\caption{Paired normal-camera and fisheye views. Acquisition conditions differ in viewpoint and resolution as well as lens distortion.}
\label{fig:paired}
\end{figure}

A RYS-4K-7730 fisheye camera was placed approximately 1 m to the right front of each speaker. Its built-in microphone captured audio at 16 kHz, synchronized with $3840\times2160$ video at 25 frames/s. A normal camera simultaneously recorded $1280\times720$ video at 25 frames/s under the same lighting and acoustic conditions. Figure~\ref{fig:paired} illustrates the paired views. The normal-camera stream offers an aligned visual reference, but its different viewpoint and resolution prevent a distortion-only interpretation of comparisons.

Individual recordings provide clean mixture references. Paired acquisition keeps speaker identity, content, and recording time shared across visual streams, enabling repeatable visual-encoder comparisons under the same acoustic interference.

\begin{figure*}[t]
\centering
\begin{minipage}[t]{0.49\textwidth}
\centering
\includegraphics[height=68mm]{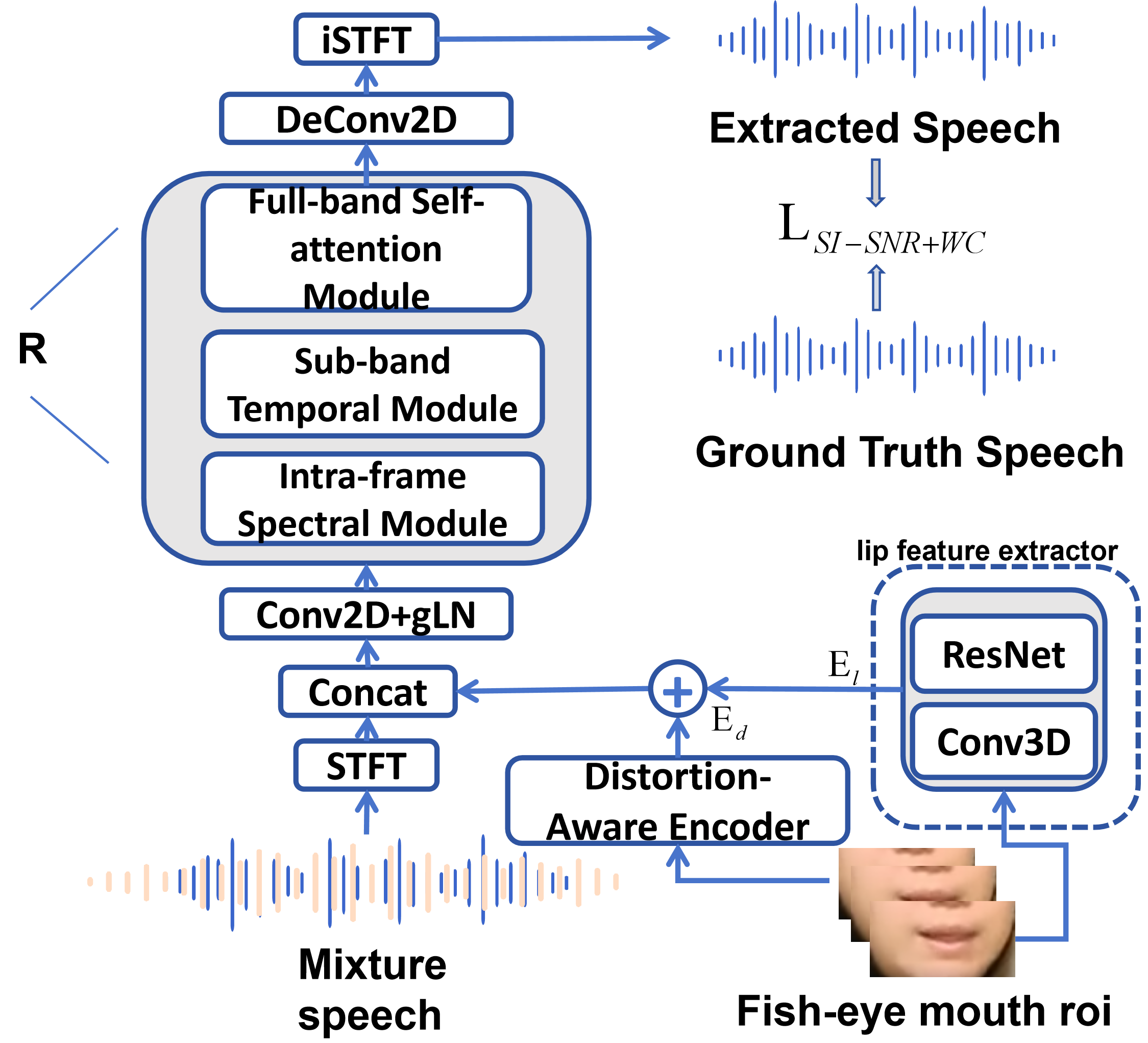}
\end{minipage}\hfill
\begin{minipage}[t]{0.49\textwidth}
\centering
\includegraphics[height=68mm]{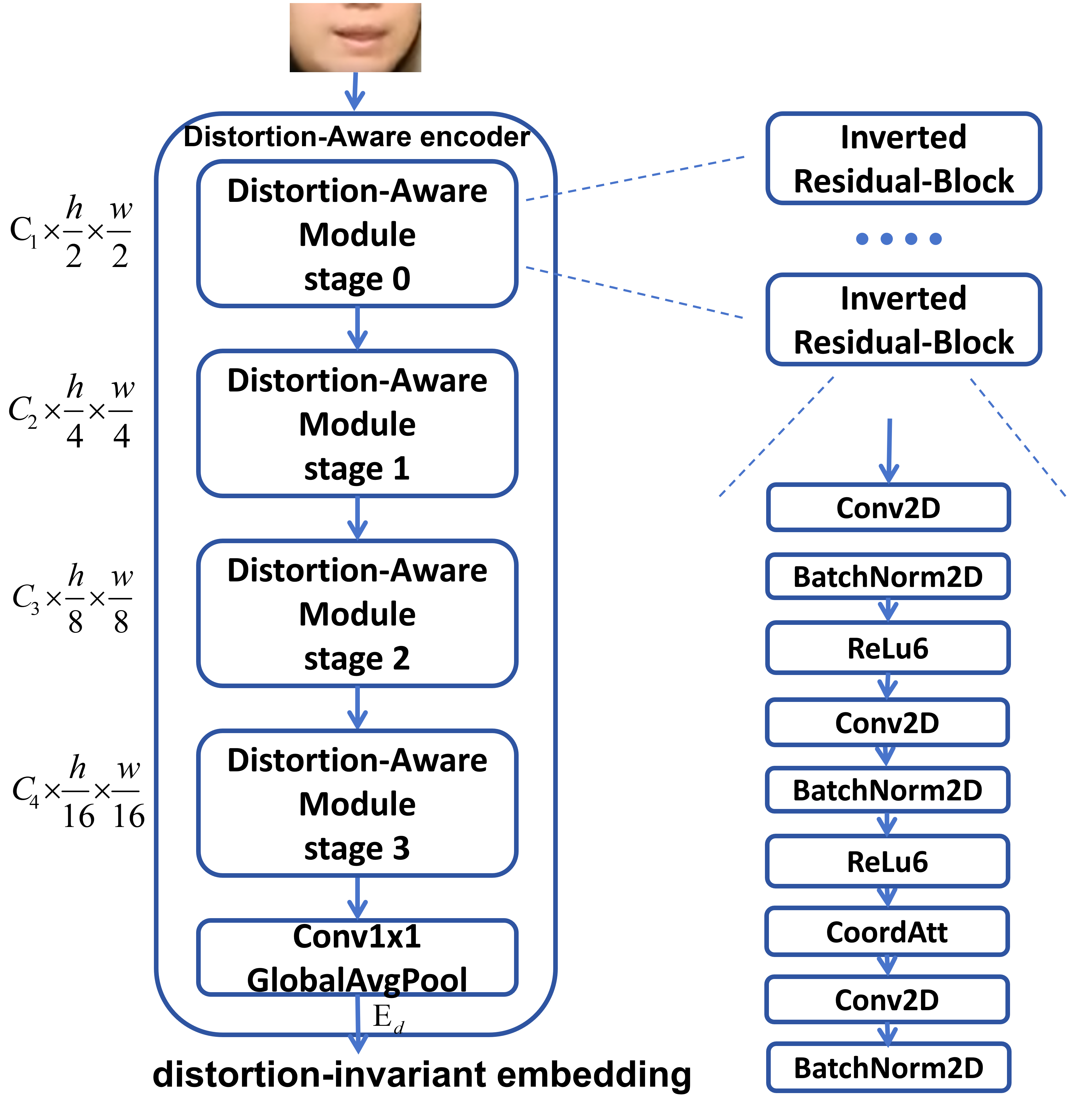}
\end{minipage}
\caption{Left: AVTSE baseline. DA features are added to lip embeddings to condition the audio separator. Right: the DA encoder has four stages of inverted residual blocks with coordinate attention; only the highest-level features are fused.}
\label{fig:model}
\end{figure*}

\subsection{Preprocessing, mixtures, and availability}
RetinaFace~\cite{deng2020retinaface} detects faces and facial landmarks. An affine transformation aligns the face to an average-face reference, after which the lip region of interest (ROI) is cropped. This alignment is part of ROI preprocessing; no camera-calibrated fisheye rectification is applied in the baseline.

The benchmark uses synthetic two-speaker mixtures of individually recorded speech. Four-second utterance segments from two different speakers are mixed at a randomly selected target-to-interferer ratio between $-10$ and $10$ dB. Training, validation, and test sets contain 15,000, 1,500, and 1,500 mixtures, respectively. Normal-camera and fisheye experiments use the same audio mixtures, speaker pairs, utterance selections, and mixing-ratio range. Thus, the benchmark permits visual-input comparisons while holding the acoustic input fixed. It evaluates simulated overlap with clean references; generalization to moving speakers, live conversations, and diverse camera placements remains to be evaluated.

The dataset is released under CC BY-NC 4.0. Dataset access, demonstrations, and baseline resources are available through the project page.\footnote{\url{https://e181d4.github.io/AISHELL8-FISHEYE/}} Mixing code, file lists, and evaluation tools are provided with the dataset release.\footnote{\url{https://huggingface.co/datasets/phot0nqqq/AISHELL8-FISHEYE}}

\section{Distortion-Aware AVTSE Baseline}
\subsection{Visual encoding}
Figure~\ref{fig:model} shows the baseline. Given a lip ROI sequence, a pretrained ResNet-18 lip-reading encoder~\cite{martinez2020lipreading} produces motion-related embeddings $E_l\in\mathbb{R}^{B\times T_v\times E}$, where $B$, $T_v$, and $E$ denote batch size, video length, and embedding dimension. We compare a frozen encoder with one updated during AVTSE training.

The DA branch is transferred from fisheye head pose estimation~\cite{zhang2025distortion} and remains frozen. It processes each frame through four stages of inverted residual blocks~\cite{sandler2018mobilenetv2} with coordinate attention~\cite{hou2021coordinate}. Only the highest-level features are fused with the lip embeddings. This branch supplies spatial representations learned on fisheye imagery; we do not assume that its output is strictly invariant to distortion.

Each inverted residual block in Figure~\ref{fig:model} (right) applies pointwise expansion, depthwise $3\times3$ filtering, coordinate attention, and pointwise projection. Convolutions use batch normalization, with ReLU6 after expansion and depthwise filtering. A residual connection is added for unit stride and matching input/output dimensions.

For a feature map $X\in\mathbb{R}^{C\times H\times W}$, coordinate attention pools along the two spatial axes:
\begin{equation}
z^h_c(i)=\frac{1}{W}\sum_{j=1}^{W}X_c(i,j),\quad
z^w_c(j)=\frac{1}{H}\sum_{i=1}^{H}X_c(i,j).
\end{equation}
The pooled features are concatenated after axis alignment and transformed by a shared $1\times1$ convolution, batch normalization, and ReLU. Splitting this representation along the spatial axis, followed by separate $1\times1$ convolutions and sigmoid activations, yields weights $g^h$ and $g^w$. The output is
\begin{equation}
Y_c(i,j)=X_c(i,j)g^h_c(i)g^w_c(j).
\end{equation}
This operation retains axis-specific positional information while aggregating spatial context. The resulting frame embeddings $E_d\in\mathbb{R}^{B\times T_v\times E}$ are added to $E_l$ to form the visual conditioning sequence.

The framewise DA and motion-related lip embeddings share a dimension for direct addition. With DA fixed, the frozen/unfrozen ResNet ablation tests whether lip adaptation is needed to exploit the transferred spatial features.

\subsection{Fusion and speech reconstruction}
The mixture short-time Fourier transform (STFT) has real and imaginary channels. We denote this tensor by $A\in\mathbb{R}^{B\times2\times F\times T_a}$; $F$ and $T_a$ are frequency bins and audio frames. The fused visual sequence $E_l+E_d$ is linearly interpolated to $T_a$ frames and projected to the frequency dimension, giving $V'\in\mathbb{R}^{B\times1\times F\times T_a}$. Concatenation produces
\begin{equation}
Z=[A;V']\in\mathbb{R}^{B\times3\times F\times T_a}.
\end{equation}
A convolutional layer and a stack of TF-GridNet blocks~\cite{wang2023tf} process $Z$. Frequency- and time-axis recurrent layers and self-attention model time-frequency dependencies. A deconvolutional layer predicts a complex mask, which is applied to the mixture STFT. The inverse STFT recovers the target waveform $\hat{s}$.

\subsection{Training objective}
We retain the original baseline's scale-aligned reconstruction objective with a waveform constraint:
\begin{equation}
\mathcal{L}=-10\log_{10}\frac{\|s\|_2^2}
{\|\alpha\hat{s}-s\|_2^2}+\|s-\hat{s}\|_1,
\quad \alpha=\frac{\hat{s}^{\top}s}{\|\hat{s}\|_2^2},
\end{equation}
where $s$ is the clean target. The first term rescales the estimate before measuring reconstruction error, while the waveform term also constrains its amplitude. Models are implemented in PyTorch and trained with Adam at an initial learning rate of $10^{-3}$ for at most 100 epochs. The learning rate is halved after three epochs without validation improvement; training stops after ten. The gradient $\ell_2$ norm is clipped to 1.

\section{Experiments and Results}
\subsection{Evaluation setup}
We evaluate AISHELL8-FISHEYE using scale-invariant signal-to-distortion ratio improvement (SI-SDRi)~\cite{le2019sdr}, short-term objective intelligibility (STOI)~\cite{stoi}, and perceptual evaluation of speech quality (PESQ)~\cite{pesq}. Higher scores are better. Ablations vary the camera stream, lip versus face ROI, ResNet-18 adaptation, and DA branch inclusion. The DA branch stays frozen whenever present.

We also evaluate the separator with a frozen lip encoder and no DA branch on LRS3~\cite{afouras2018lrs3}. This experiment uses 16-kHz, three-second audio segments, two different speakers per mixture, and mixing ratios from $-5$ to $5$ dB. There are 50,000 training, 5,000 validation, and 3,000 test mixtures. Published comparison scores in Table~\ref{tab:lrs3} are taken from~\cite{li2024audio,li2023iianet}; they provide context rather than a uniformly retrained comparison under identical mixture lists.

\begin{table}[t]
\centering
\caption{LRS3 results. Comparison scores are reported in~\cite{li2024audio,li2023iianet}. SI-SDRi and SDRi are in dB.}
\label{tab:lrs3}
\begin{tabular}{lrrr}
\toprule
Method & SI-SDRi & SDRi & PESQ\\
\midrule
AVConvTasNet~\cite{wu2019time} & 11.2 & 11.7 & 2.58\\
VisualVoice~\cite{gao2021visualvoice} & 9.9 & 10.3 & 2.13\\
AVLiT-8~\cite{martel2023audio} & 13.5 & 13.6 & 2.78\\
CTCNet~\cite{li2024audio} & 17.4 & 17.5 & 3.24\\
IIANet~\cite{li2023iianet} & 18.3 & 18.5 & 3.28\\
Baseline (no DA) & 17.9 & 17.9 & 3.22\\
\bottomrule
\end{tabular}
\end{table}

\subsection{Camera comparison and visual adaptation}
The baseline obtains 17.9 dB SI-SDRi on LRS3, compared with the reported 17.4 dB for CTCNet and 18.3 dB for IIANet. This result supports its use as an AVTSE baseline without establishing superiority over existing methods.

Table~\ref{tab:ablation} reports all AISHELL8-FISHEYE configurations. With a frozen lip encoder and no DA branch, normal-camera video yields 19.33 dB SI-SDRi versus 10.01 dB for fisheye video, a 9.32 dB gap. The strongest evaluated fisheye configuration reaches 11.84 dB, leaving a 7.49 dB gap to the normal-camera baseline. These results show the difficulty of the fisheye acquisition setting, but differences in viewpoint and resolution also contribute possible confounds.

\begin{table}[t]
\centering
\caption{AISHELL8-FISHEYE ablations. F/T denotes a frozen/trainable lip ResNet; DA denotes the distortion-aware branch. SI-SDRi is in dB.}
\label{tab:ablation}
\setlength{\tabcolsep}{5pt}
\begin{tabular}{llccrrr}
\toprule
Video & ROI & Res. & DA & SI-SDRi & STOI & PESQ\\
\midrule
Normal & Lip & F & -- & 19.33 & 0.90 & 3.41\\
Normal & Lip & F & + & 19.01 & 0.89 & 3.37\\
\midrule
Fisheye & Lip & F & + & 9.91 & 0.77 & 2.90\\
Fisheye & Lip & T & + & \textbf{11.84} & \textbf{0.81} & \textbf{2.96}\\
Fisheye & Lip & F & -- & 10.01 & 0.77 & 2.92\\
Fisheye & Lip & T & -- & 10.47 & 0.78 & 2.93\\
Fisheye & Face & T & -- & 10.30 & 0.78 & 2.74\\
\bottomrule
\end{tabular}
\end{table}

The DA branch helps only in combination with visual adaptation in the tested configurations. With a trainable lip encoder, adding DA improves SI-SDRi by 1.37 dB, from 10.47 to 11.84 dB; STOI rises from 0.78 to 0.81 and PESQ from 2.93 to 2.96. With a frozen lip encoder, DA instead changes SI-SDRi from 10.01 to 9.91 dB. For normal video, it changes 19.33 to 19.01 dB. Thus, the results do not support a universal benefit from DA features. These scores are point estimates without repeated-run uncertainty or significance tests.

Unfreezing the lip encoder improves fisheye SI-SDRi by 1.93 dB with DA and by 0.46 dB without it. This interaction suggests that adapting the lip representation helps the separator use the transferred spatial features. Without DA, lip ROIs slightly outperform face ROIs in SI-SDRi (10.47 versus 10.30 dB) and PESQ (2.93 versus 2.74), while STOI is unchanged. These observations favor the adapted lip-plus-DA configuration among the evaluated variants.

The three metrics give a consistent direction for the DA comparison with a trainable lip encoder: signal fidelity, intelligibility, and perceptual quality all improve. However, the gain remains much smaller than the gap to normal-camera guidance. The baseline should therefore be viewed as evidence that complementary visual features can help in this setting, while leaving substantial room for better distortion handling. The face-versus-lip comparison is narrower: it covers only the trainable configuration without DA and does not establish that lip crops are preferable for every architecture or training strategy.

Further comparisons against rectification, synthetic distortion augmentation, and a matched generic attention branch would help distinguish the effects of fisheye pretraining from additional visual capacity. These alternatives are not evaluated in the present benchmark.

\section{Conclusion}
AISHELL8-FISHEYE provides 47.03 hours of paired fisheye and normal-camera recordings with speaker-disjoint AVTSE evaluation. A frozen DA branch combined with an adapted lip encoder improves fisheye SI-SDRi from 10.47 to 11.84 dB, but does not close the gap to normal-camera inputs. The dataset and baseline enable reproducible study of speech extraction under distorted visual acquisition and expose the importance of jointly adapting lip features.

\clearpage
\section{Compliance with Ethical Standards}
All recruited participants signed consent forms covering data collection and public release. AISHELL8-FISHEYE is released under CC BY-NC 4.0.

\section{Generative AI Use Disclosure}
Generative AI tools were used for limited language editing purposes, including improving clarity and correcting grammatical issues. No substantive content, analysis, or conclusions were generated by AI. The authors remain fully responsible for the content of this manuscript.
\bibliographystyle{IEEEbib}
\bibliography{sample-base}
\end{document}